\documentclass[iop]{emulateapj}
\usepackage{apjfonts}
\usepackage{graphicx}
\usepackage{color}

\def\d3{$\delta_{3}$ }
\def\1d3{$(1 + \delta_{3})$ }
\def\l1d3{$\log_{10}(1 + \delta_{3})$ }

\def\s3{$\Sigma_{3}$}
\def\ha{H$\alpha$}
\def\hb{H$\beta$}

\def\24m{24 $\mu$m}

\def\sm{$\rm~M_{*}$}

\def\kms{${\rm km~s^{-1}}$ }

\def\Msolar{$\rm M_{\odot}$}

\def\rmxaa{RMxAA}

\def\sigsm{$\Sigma_{*}$}
\def\sigSFR{$\Sigma_{\rm~SFR}$}
\def\h2{$\rm~H_{2}$}
\def\Mh2{$\rm~M_{H2}$}
\def\sigh2{$\Sigma_{\rm~H_{2}}$}
\def\fgas{$f_{\rm~gas}$}

\shorttitle{Green Valley Galaxies}
\shortauthors{Lin et al.}

\begin{document}

\title{Resolved star formation and molecular gas properties of green valley galaxies: a first look with ALMA and MaNGA}


\author{Lihwai Lin \altaffilmark{1}, Francesco Belfiore \altaffilmark{2,3,10}, Hsi-An Pan \altaffilmark{1}, M. S. Bothwell \altaffilmark{2,3}, Pei-Ying Hsieh \altaffilmark{1}, Shan Huang \altaffilmark{4}, Ting Xiao \altaffilmark{5}, Sebasti\'{a}n F. S\'{a}nchez \altaffilmark{6}, Bau-Ching Hsieh \altaffilmark{1}, Karen Masters \altaffilmark{7}, S. Ramya \altaffilmark{5}, Jing-Hua Lin \altaffilmark{1,8}, Chin-Hao Hsu \altaffilmark{1,8}, Cheng Li \altaffilmark{9}, Roberto Maiolino \altaffilmark{2,3}, Kevin Bundy \altaffilmark{10}, Dmitry Bizyaev \altaffilmark{11,12}, Niv Drory \altaffilmark{13}, H\'{e}ctor Ibarra-Medel \altaffilmark{6},  Ivan Lacerna \altaffilmark{14,15,16}, Tim Haines \altaffilmark{17}, Rebecca Smethurst \altaffilmark{18}, David V. Stark \altaffilmark{19}, Daniel Thomas \altaffilmark{7}}

\altaffiltext{1}{Institute of Astronomy \& Astrophysics, Academia Sinica, Taipei 10617, Taiwan; Email: lihwailin@asiaa.sinica.edu.tw}
\altaffiltext{2}{Cavendish Laboratory, University of Cambridge, 19 J. J. Thomson Avenue, Cambridge CB3 0HE, United Kingdom}
\altaffiltext{3}{University of Cambridge, Kavli Institute for Cosmology, Cambridge, CB3 0HE, UK.}
\altaffiltext{4}{Center for Cosmology and Particle Physics, New York University, New York, NY 10003, USA}
\altaffiltext{5}{Shanghai Astronomical Observatory, Chinese Academy of Science, 80 Nandan Road, Shanghai 200030, China}
\altaffiltext{6}{Instituto de Astronom\'{i}a, Universidad Nacional Auton\'{o}ma de Mexico, A.P. 70-264, 04510, M\'{e}xico, D.F., M\'{e}xico}
\altaffiltext{7}{Institute of Cosmology \& Gravitation, University of Portsmouth, Dennis Sciama Building, Portsmouth, PO1 3FX, UK}
\altaffiltext{8}{Department of Physics, National Taiwan University, 10617, Taipei, Taiwan}

\altaffiltext{9}{Tsinghua Center of Astrophysics \& Department of Physics, Tsinghua University, Beijing 100084, China}
\altaffiltext{10}{UCO/Lick Observatory, University of California, Santa Cruz, 1156 High St. Santa Cruz, CA 95064, USA}

\altaffiltext{11}{Apache Point Observatory and New Mexico State
University, P.O. Box 59, Sunspot, NM, 88349-0059, USA}

\altaffiltext{12}{Sternberg Astronomical Institute, Moscow State
University, Moscow, Russia}

\altaffiltext{13}{McDonald Observatory, University of Texas at Austin, 1
University Station, Austin, TX 78712-0259, USA}

\altaffiltext{14}{Instituto Milenio de Astrof\'isica, Av. Vicu\~na Mackenna 4860, Macul, Santiago, Chile}

\altaffiltext{15}{Instituto de Astrof\'isica, Pontificia Universidad Cat\'olica de Chile, Av. Vicuna Mackenna 4860, 782-0436 Macul, Santiago, Chile}

\altaffiltext{16}{Astrophysical Research Consortium, Physics/Astronomy Building, Rm C319, 3910 15th Avenue NE, Seattle, WA 98195, USA}   

\altaffiltext{17}{Department of Astronomy, University of Wisconsin-Madison, 475N. Charter St., Madison WI 53703, USA}

\altaffiltext{18}{School of Physics and Astronomy, University of Nottingham, University Park, Nottingham, NG7 2RD, UK	}

\altaffiltext{19}{Kavli Institute for the Physics and Mathematics of the Universe (WPI), The University of Tokyo Institutes for Advanced Study, The University of Tokyo, Kashiwa, Chiba 277-8583, Japan}

\begin{abstract}

We study the role of cold gas in quenching star formation in the green valley by analysing ALMA $^{12}$CO (1-0) observations of three galaxies with resolved optical spectroscopy from the MaNGA survey. 
We present resolution-matched maps of the star formation rate and molecular gas mass. These data are used to calculate the star formation efficiency (SFE) and gas fraction  (\fgas) for these galaxies separately in the central `bulge' regions and outer disks. We find that, for the two galaxies whose global specific star formation rate (sSFR) deviates most from the star formation main sequence, the gas fraction in the bulges is significantly lower than that in their disks, supporting an `inside-out' model of galaxy quenching. For the two galaxies where SFE can be reliably determined in the central regions, the bulges and disks share similar SFEs. This suggests that a decline in \fgas~ is the main driver of lowered sSFR in bulges compared to disks in green valley galaxies. Within the disks, there exist common correlations between the sSFR and SFE and between sSFR and \fgas~on kpc scales -- the local SFE or \fgas~in the disks declines with local sSFR. Our results support a picture in which the sSFR in bulges is primarily controlled by \fgas, whereas both SFE and \fgas~ play a role in lowering the sSFR in disks. A larger sample is required to confirm if the trend established in this work is representative of green valley as a whole.

\end{abstract}

\keywords{galaxies:evolution $-$ galaxies: low-redshift $-$}

\section{INTRODUCTION}

It has been known for more than a decade that the distributions of galaxy properties are bimodal in either the color-magnitude diagram (CMD) or the relation between the star formation rate (SFR) and stellar mass \citep{bla03,kau03,bal04}. Galaxies in between the blue cloud and the red sequence, the so-called `green valley' galaxies \citep[see][for a review on this topic]{sal14}, are often thought to be in transition from the star-forming phase to the quiescent phase \citep{bel04,fab07,mar07}. Under this framework, the scarce density of green valley galaxies implies that either the fraction of star-forming galaxies undergoing star formation quenching is low or the quenching process is fast enough so that the lifetime in the green valley phase is short. Properties of these green valley galaxies thus carry important information on how the star formation is quenched. 

A study carried out by \citet{sch14} has shown that the color-selected transitional galaxies are dominated by galaxies with late-type (disk) morphology with a slowly-declining star formation history, rather than morphologically early-type (elliptical) galaxies of which the star formation is shut down abruptly \citep[also see][]{sme15}.  However, it remains unclear what physical mechanism plays the dominant role in suppressing the star formation of galaxies and produces these two types of green valley galaxies. Furthermore, the underlying star formation history of green valley galaxies can be even more complicated if galaxies are rejuvenated by accreting fresh gas through minor mergers \citep{hai15,lac16}.

A variety of scenarios have been proposed to explain the shutdown of star formation in galaxies, usually split into so-called `nature' processes - referring to the consequences of internal evolution of galaxies, and `nurture', or the impact of the environment a galaxy lives in. If `nature' processes dominate, galaxies would be expected to grow, evolve, and die inside-out \citep{whi91,mo98}. Recent IFU observations demonstrate that many nearby spiral galaxies show negative gradients in stellar ages and metallicities, supporting this inside-out picture \citep{san14,gon14,li15,god17,bel17}. On the other hand, if `nurture' dominates galaxy evolution, external processes such as ram-pressure stripping \citep{gun72}, high speed galaxy encounters \citep{moo96}, galaxy mergers \citep{mih94}, and `strangulation \citep{lar80,bal00,pen15}, are responsible for quenching. In this picture, star formation quenching is likely to occur globally or in the outer regions of galaxies first due to the lack of continuous supply for the cold gas reservoir.

Previous works in the area of green valley galaxies faced two main limitations. Firstly, earlier optical studies on
transitional galaxies largely rely on the single-fiber SDSS spectroscopy, which lacks spatial information and covers only the central part of nearby galaxies. Secondly, although the star formation histories (including recent and on-going SFR) of galaxies can be inferred from UV, optical to infrared data based on the broadband SEDs (spectral energy distributions) and spectral lines, a complete picture of the galaxy evolution processes requires understanding of the cold molecular gas, which serves as the fuel of star formation. In this work, we present the ALMA CO observations of three green valley galaxies selected from the SDSS-IV Mapping Nearby Galaxies
at Apache Point Observatory \citep[MaNGA;][]{bun15,law16,yan16a,yan16b}. The focuses of this work are to characterize the role of cold molecular gas in the star formation quenching and to probe the sequence of quenching among substructures of galaxies (e.g., bulge vs. disk) by combining spatially resolved observations of the stellar population and molecular gas. Specifically, we will address whether the declining star formation activity is caused by a depletion of gas or by a suppression of star-forming efficiency in different galactic regions.

Throughout this paper we adopt the following cosmology: \textit{H}$_0$ = 100$h$~\kms Mpc$^{-1}$, $\Omega_{\rm m} = 0.3$ and $\Omega_{\Lambda } = 0.7$. We use a Salpeter IMF and adopt the Hubble constant $h$ = 0.7. All magnitudes are given in the AB system.

\section{DATA \label{sec:data}}
\subsection{MaNGA Targets}

\begin{deluxetable*}{cccccccccccccc}
\tablecaption{Properties of the three MaNGA galaxies.\label{tab:property}}
\tablehead{\colhead{ID} & \colhead{MaNGA ID} & \colhead{RA} & \colhead{Dec} & \colhead{Redshift}  & \colhead{$\log $(M$_\star$/\Msolar)} &  \colhead{$\log$ ($\frac{\rm SFR}{\rm M_{\odot} yr^{-1}}$)} & $\Delta$sSFR & \colhead{$\log $(M$_{\rm H2}$/\Msolar)} & \colhead{$\log $(M$_{\rm HI}$/\Msolar)} &  \colhead{$\log$ ($\frac{\rm SFE}{\rm yr^{-1}}$)$^{(b)}$}\\
&&&&(D$_{r}^{(a)}$)}
\startdata
1 & 1-596678 & 332.89284 & 11.79593 & 0.02695 &  10.88 & 0.46 & $-$0.24 & 9.47$^{(c)}$ & 10.21 & -9.0 \\
& & & &(114.7 Mpc)\\
2 & 1-114956 & 332.79873 & 11.80073 & 0.02702 &  10.36 & $-$0.3 & $-$0.48 & 8.98$^{(d)}$ & 9.87 & -9.2\\
& & & &(115.0 Mpc)\\
3 & 1-596598 & 331.12290 & 12.44263 & 0.02659 &  10.98 & 0.075 & $-$0.73 & \nodata & 9.7$^{(e)}$ & \nodata \\
& & & &(113.2 Mpc)
\enddata
\tablecomments{$^{(a)}$Comoving radial distance; $^{(b)}$The global SFE estimated based in the single-dish CO measurements; $^{(c)}$Data taken with JCMT by Ting Xiao et al.; $^{(d)}$Data taken from \citet{sai12}; $^{(e)}$Galaxy 1 is in the edge of the GBT beam (5.5\arcmin away) of Galaxy 3 and both galaxies are at very similar redshifts, so the HI can be attenuated flux from the HI linked to that in the edges of the beam. If there is no HI detected linked to Galaxy 3, the upper limit (assuming a width of 400 km s$^{-1}$) would be $\log $(M$_{\rm HI}$/\Msolar) = 9.22 instead.}
\end{deluxetable*}

 MaNGA is an on-going integral field unit (IFU) survey on the SDSS 2.5m telescope (Gunn et al. 2006), as part of the SDSS-IV survey \citep{alb17,bla17}. MaNGA makes use of a modification of the BOSS spectrographs (Smee et al. 2013) to bundle fibres into hexagons (Drory et al. 2015). Each spectra has a wavelength coverage of 3500-10,000\AA,  and instrumental resolution $\sim$60 kms$^{-1}$ . After dithering, MaNGA data have an effective spatial resolution of 2.5\arcsec (FWHM; Law et al. 2015), and data cubes are gridded with 0.5\arcsec spaxels. 
 
 We make use of the Pipe3D pipeline \citep{san16a} to model the stellar continuum with 156 templates with 39 ages and 4 stellar populations that were extracted from  a combination of the synthetic stellar spectra from the GRANADA library \citep{mar05} and the MILES project \citep{san06,vaz10,fal11}. Details of the fitting procedures are described in \citet{san16b}. In short, a spatial binning is first performed in order to reach a S/N of
50 accross the entire field of view (FoV) for each datacube. A stellar
population fit of the coadded spectra within each spatial bin is then
computed. 
The stellar population model for spaxels with continuum S/N $>$ 3 is then estimated by
re-scaling the best fitted model within each spatial bin to the continuum
flux intensity in the corresponding spaxel, following \citet{cid13} and \citet{san16a}.  The stellar mass surface density (\sigsm) is then obtained using the stellar mass derived for each spaxel and then normalized to the physical area of one spaxel. We derive the emission line fluxes following the same procedure described in \citet{bel16}. Briefly speaking, the fitting are performed on continuum subtracted spectra using sets of Gaussians (one per line) with a common velocity. The dust attenuation is corrected by using the Balmer decrement, adopting the \citet{cal01} attenuation curve with $Rv = 4.05$ and a theoretical value for the Balmer line ratio (\ha/\hb = 2.86) taken from \citet{ost06}, assuming case B recombination. SFR is then estimated based on this extinction corrected \ha~flux using the conversion given by \citet{ken98a} with the Salpeter IMF. Similarly, we convert the spaxel-based SFR into the SFR surface density (\sigSFR) by normalizing it to the spaxel area. At at fixed extinction curve and IMF, the uncertainty in the SFR estimate is proportional to that of the \ha~flux and is less than 33\% given that we only limit to our analysis to spaxels with S/N (\ha) > 3.

We show in the left panel of Figure \ref{fig:sfr-sm}, the locations of the sSFR, defined as the SFR divided by the stellar mass (\sm), versus \sm~for 2730 MaNGA galaxies (black dots), from an internal release (labeled MPL5), very closely equivalent to Data release 13 (Albareti et al 2017), by integrating the Pipe3D results from individual MaNGA spaxels.  
The three green valley targets (MaNGA 1-596678, 1-114956, and 1-596598) for the ALMA follow-up, highlighted by the color-coded stars, were drawn from the first 118 galaxies observed by MaNGA at the time when the ALMA proposal was prepared. They are randomly selected to be massive galaxies that lie below the star-forming main sequence relation with different separations from the main sequence, $\Delta$sSFR, defined as the offset in log(sSFR) relative to the main sequence value (i.e., log(sSFR) - log(sSFR$_{MS}$)). Previous studies have revealed significant differences in the slope and normalization of the main sequence \citep[e.g., see][]{spe14}. The selection of the star-forming population, the method determining the star formation rate and stellar mass, as well as the IMF, have a strong effect in determining the properties of the main sequence. In light of this complexity, we compute our own value of sSFR$_{MS}$ based on the Pipe3D results to be self-consistent. The sSFR of the main sequence is determined to be $\sim$ 10$^{-10.18}$ yr$^{-1}$, as shown in the blue solid line, by taking the the median sSFR of galaxies with log(sSFR/yr$^{-1}$) > -10.6. Our derived sSFR$_{MS}$ is close to the $z \sim 0$ value (sSFR $\sim$ 10$^{-10.09}$ yr$^{-1}$) derived using the empirical sSFR vs. redshift relation given in Equation (13) of \citet{elb11}. We also require the targets to be accessible by ALMA and we do not impose the constraint on the predicted CO abundance when selecting the targets. We number them 1 to 3 (hereafter Galaxy 1, Galaxy 2, and Galaxy 3) according to their $\Delta$sSFR(see Table 1). Although Galaxy 1 lies close to the lower edge of the star-forming main sequence on the global sSFR -- \sm~ plane, all three galaxies are referred to as `green valley galaxies' loosely in this work.

These three objects were recently observed as part of the HI-MaNGA programme at the Robert C. Byrd Green Bank Telescope (GBT), which is obtaining HI 21cm observations of a large sample of MaNGA galaxies (AGBT17A\_012, PI: K. Masters). Galaxies 1 and 2 have HI gas fractions comparable to that of the normal HI galaxies while Galaxy 3 is below the ALFALFA scaling relation \citep[see Figure 2(c) of][]{hua12}. In addition, Galaxy 1 and Galaxy 2 were also observed in CO (2-1) with JCMT (PI: Ting Xiao) and CO (1-0) with IRAM \citep{sai12}, respectively, from which the total H$_{2}$ mass can be derived. The beam size of JCMT is 22$\arcsec$ and is 32.5$\arcsec$ for IRAM. The general properties of the three green valley galaxies are summarized in Table 1. We assume the CO(2-1) to CO(1-0) ratio to be 0.7 when calculating the total H$_{2}$ mass of Galaxy 1.

\subsection{ALMA Observations}

Molecular gas observations in $^{12}$CO(1-0) were carried out with ALMA in Cycle 3 on January 2016 using Band 3 receiver (project code: 2015.1.01225.S; PI: Lihwai Lin). The baseline ranges from 15 to 310 meters. The largest structure that we expect to be sensitive to is about 36$\arcsec$ ($\sim$ 20 kpc). Thus, the missing flux should be negligible.  Uranus was observed as flux calibrator for Galaxy 2, and Neptune was used for Galaxy 1 and Galaxy 3. The phase and bandpass of the observations of Galaxy 1 and Galaxy 2 were calibrated with J2232+1143 and J2222+1213 , respectively, and J2200+1030  and J2148+0657  for Galaxy 3. The on-source time is $\sim$ 1 hr for each galaxy.

Our spectral setup includes one line targeting $^{12}$CO (1-0). The window has a bandwidth of 0.937 GHz (2500 km s$^{-1}$), with a channel width of 3906.250 kHz (10.1 km s$^{-1}$). The data were processed by pipeline (version r35932 and r36660) in the Common Astronomy Software Applications package (CASA, version 4.5.1 r35996 and 4.5.3 r36115). 

The task CLEAN was employed for deconvolution with a robust = 0.5 weighting (Briggs).
We adopted a user-specified image center, pixel size, and restoring beamsize to match the image grid and the spatial resolution of the MaNGA images during the CLEAN process.
The user-specified image center is $\sim$ 0.1$\arcsec$ away from the original center in the ALMA observations. We adopt a geometric mean beamsize of the user-specified beam,  2.5$\arcsec$ $\times$ 2.5$\arcsec$ ($\sim$ 1.4 $\times$ 1.4 kpc), similar to that of the native beamsize reported by the CLEAN (2.6$\arcsec$ $\times$ 2.2$\arcsec$).
We have confirmed that all results remain unchanged if we instead use the original image center and restoring beamsize.
Sensitivity of the three observations are almost identical.   The final cubes have  channel width of 10.1 km s$^{-1}$ and rms noise ($\sigma_{\mathrm{rms}}$) of $\sim$ 0.5 mJy beam$^{-1}$. Integrated intensity maps were created from the cubes with a clip in noise of 1.5-$\sigma_{\mathrm{rms}}$. Varying the clipping threshold from 2- to 1.3-$\sigma$ results in a change of the CO flux from -15\% to +10\% with respect to the case using 1.5-$\sigma$. Since the ALMA observations have larger field of view than MaNGA, the edge of ALMA maps were cut off to match the image size of MaNGA.  The H$_{2}$ mass surface density (\sigh2) is computed from the CO surface density by adopting a conversion factor ($\alpha_{\mathrm{CO}}$) of 4.3 \Msolar (K km s$^{-1}$ pc$^{2}$)$^{-1}$ \citep[e.g.,][]{bol13}.

We can compare the total CO flux and/or H$_{2}$ mass obtained by integrating the ALMA results with those based on single-dish observations for two of our targets. We find that the total ALMA CO(1-0) flux for Galaxy 1 is in good agreement with that derived from the JCMT CO(2-1) observation if adopting a conventional CO(2-1)/CO(1-0) ratio = 0.7. For Galaxy 2, which is part of the COLD GASS sample, its ALMA-integrated H$_{2}$ mass is factor of 1.9 lower than the value listed in the COLD GASS catalog when applying the same $\alpha_{\mathrm{CO}}$. We suspect that this discrepancy may be related to the method of aperture correction used in the COLD GASS estimation for this object.

\section{Results}

\begin{figure*}
\centering
\includegraphics[angle=0,width=0.9\textwidth]{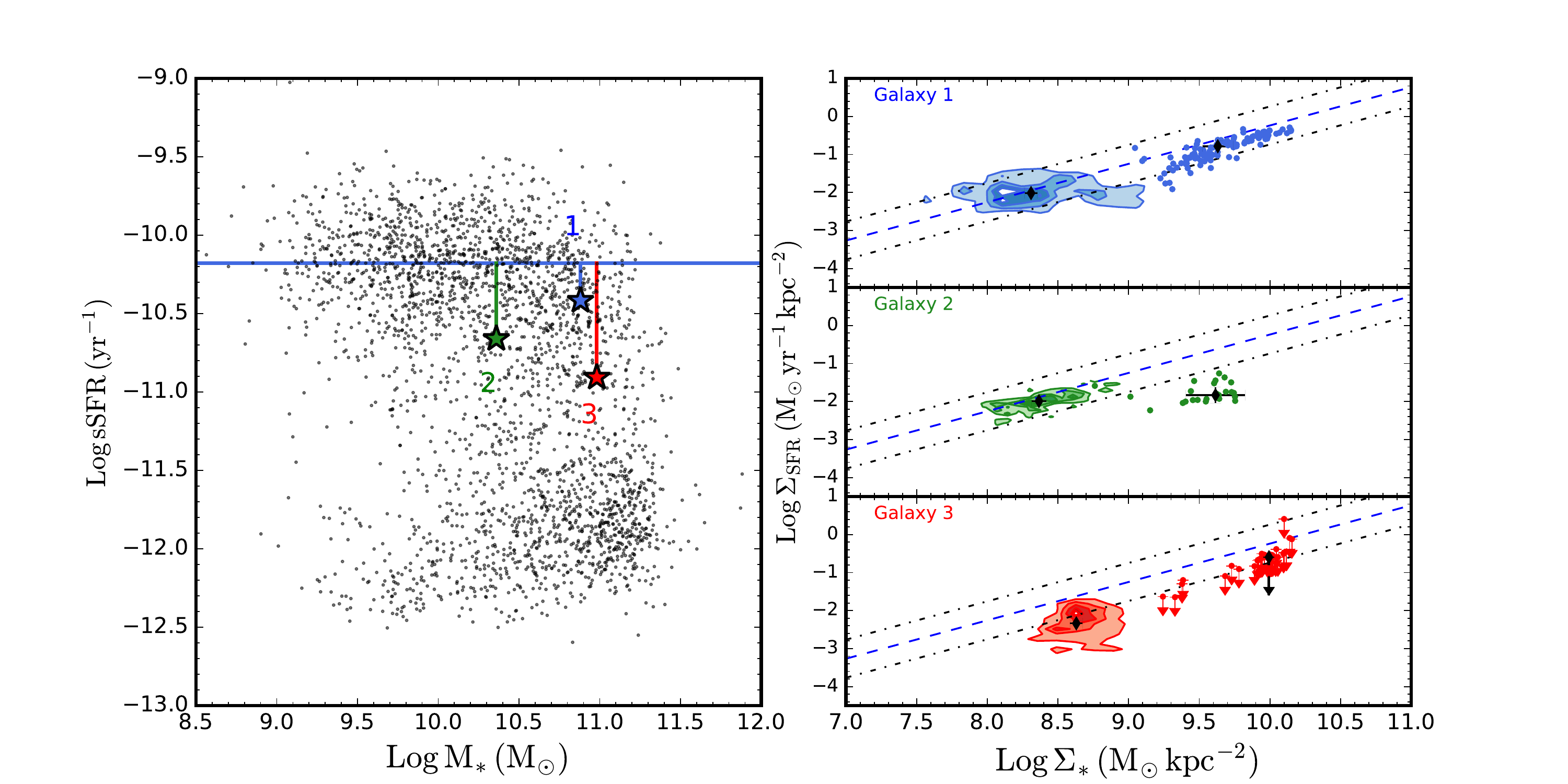}
\caption{Left panels: The positions of the three green valley galaxies on the global sSFR versus stellar mass plane derived from the Pipe3D analysis of MaNGA data. Each MaNGA MPL5 galaxy is shown as a black dot. The large stars show the 3 green valley galaxies: Galaxy 1 (blue), Galaxy 2 (green), and Galaxy 3 (red).  The horizontal line represents a constant log(sSFR/yr$^{-1}$)  = -10.18, denoting the typical value of the star-forming population. Right panels: The resolved SFR surface density versus stellar mass surface density of the 3 green valley galaxies (from top to bottom: Galaxy 1, Galaxy 2, and Galaxy 3). MaNGA 0.5\arcsec spaxels belonging to the bulges are shown as solid circles (note that these are not all independent data point as the effective resolution of MaNGA is 2.5\arcsec) while the contours show the distributions of the spaxels in the disks. The black diamond symbols denote the median values for both disk and bulge components. The blue dashed line represents the best-fit of the resolved relation for the main-sequence galaxies \citep{hsi17}, corresponding to sSFR = 10$^{-10.33}$ yr$^{-1}$ while the upper and lower dot-dashed lines show sSFR = 10$^{-9.83}$ and 10$^{-10.83}$ yr$^{-1}$, respectively. SFR for the bulge of Galaxy 3 are shown as upper limits due to possible AGN contaminations. \label{fig:sfr-sm}.}
\end{figure*}

\begin{figure*}
\centering
\includegraphics[angle=0,width=0.85\textwidth]{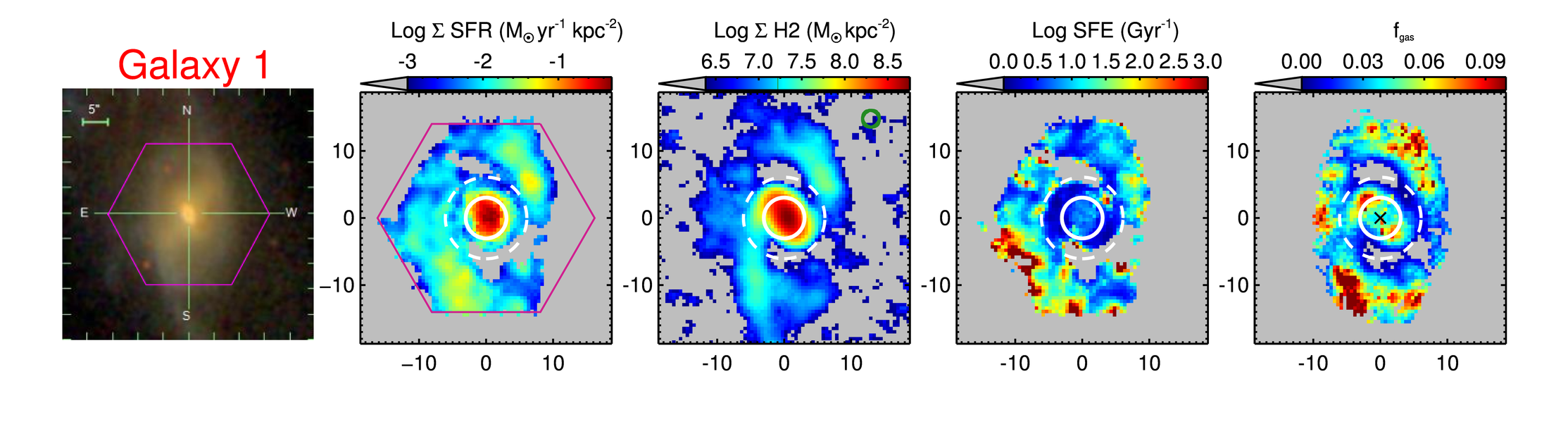}\\
\includegraphics[angle=0,width=0.85\textwidth]{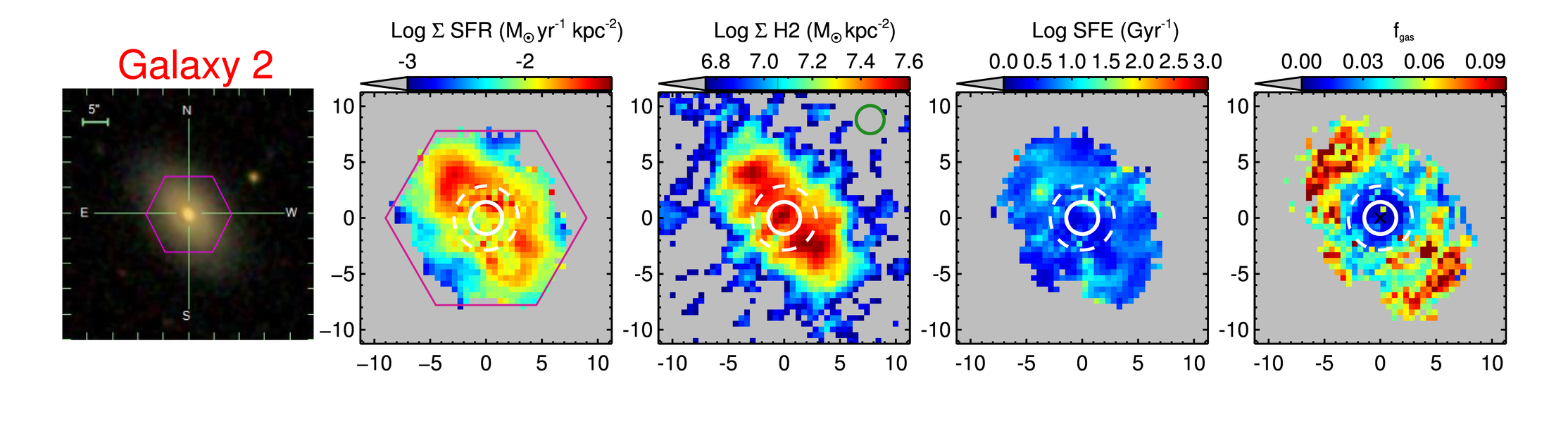}\\
\includegraphics[angle=0,width=0.85\textwidth]{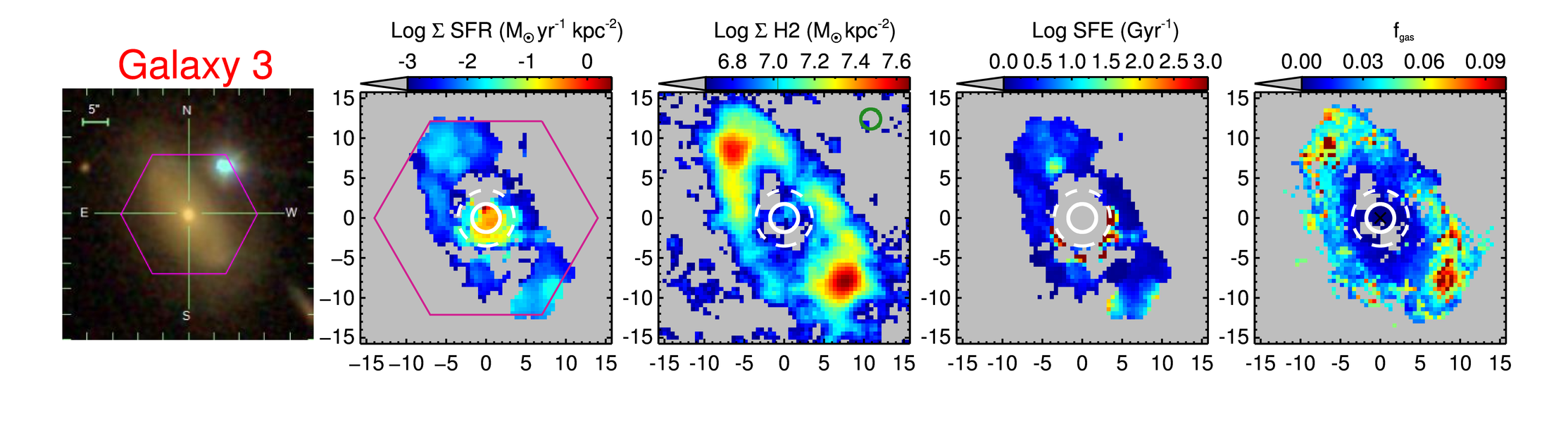}
\caption{Maps of various quantities for the 3 green valley galaxies (from top to bottom: Galaxy 1, Galaxy 2, and Galaxy 3). The 1st column shows the SDSS $gri$ composite images with the MaNGA hexagon overlaid in pink. The 2nd column displays \sigSFR~ based on the MaNGA \ha~observations. The \sigSFR~ in the central region of Galaxy 3 must be interpreted as an upper limit as its  \ha~emission is likely contaminated by AGN contributions. The 3rd column shows the \h2~mass surface density map based on ALMA CO(1-0) observations. The 4th and last columns show the distributions of SFE and \fgas, respectively. In the 2nd to 5th columns, the `bulge' and `disk' regions are defined as those spaxels within the white solid circles and outside the dashed circles, respectively. \label{fig:map}}
\end{figure*}

\begin{figure*}
\centering
\includegraphics[angle=0,width=0.9\textwidth]{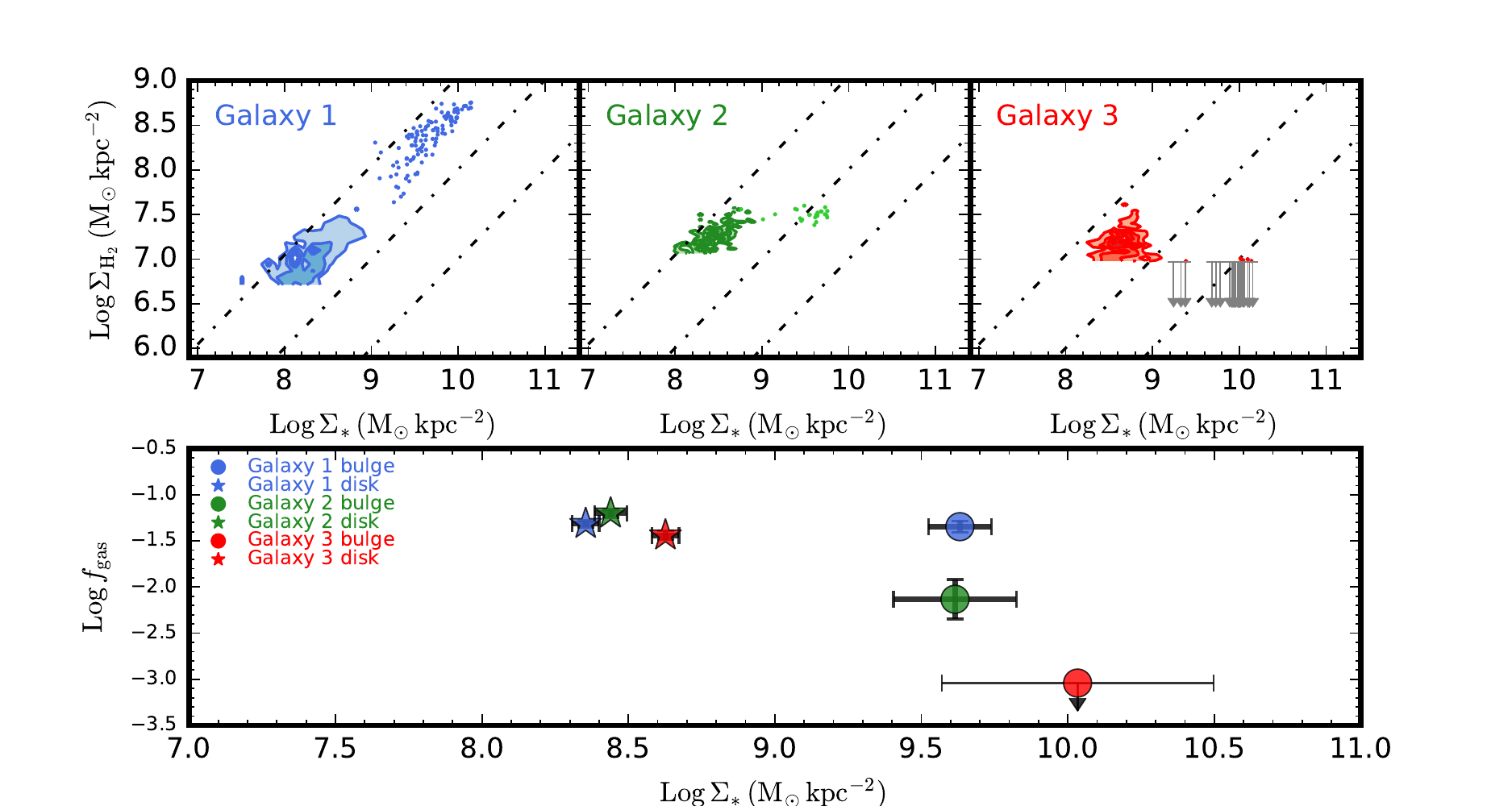}
\caption{Top panels: The gas surface density versus the stellar mass surface density relation for Galaxy 1, Galaxy 2, and Galaxy 3 (from left to right). Spaxels belonging to the bulges are shown as small dots while the contours show the distributions of the spaxels in the disks. The sharp boundary in \sigh2~corresponds to the S/N = 2 cutoff in the CO flux density. The gray arrows denote bulge spaxels falling below the detection limit. The three dashed-dotted lines correspond to constant gas fractions of 0.1, 0.01, and 0.001 (from top to bottom). Bottom panel: The gas fraction as a function of the stellar mass surface density. The median values for the bulges and disks are shown as circles and stars, respectively. The symbols are color-coded to represent different galaxies (blue: Galaxy 1; green: Galaxy 2; red: Galaxy 3). The errors bars denote the uncertainties in the median values, calculated as the standard deviation normalized by the square root of the number of independent spaxels. In some cases, the error bars are smaller than the size of the symbols and hence are invisible from the plots.  \label{fig:gas-sm}}
\end{figure*}

\begin{figure*}
\centering
\includegraphics[width=0.9\textwidth]{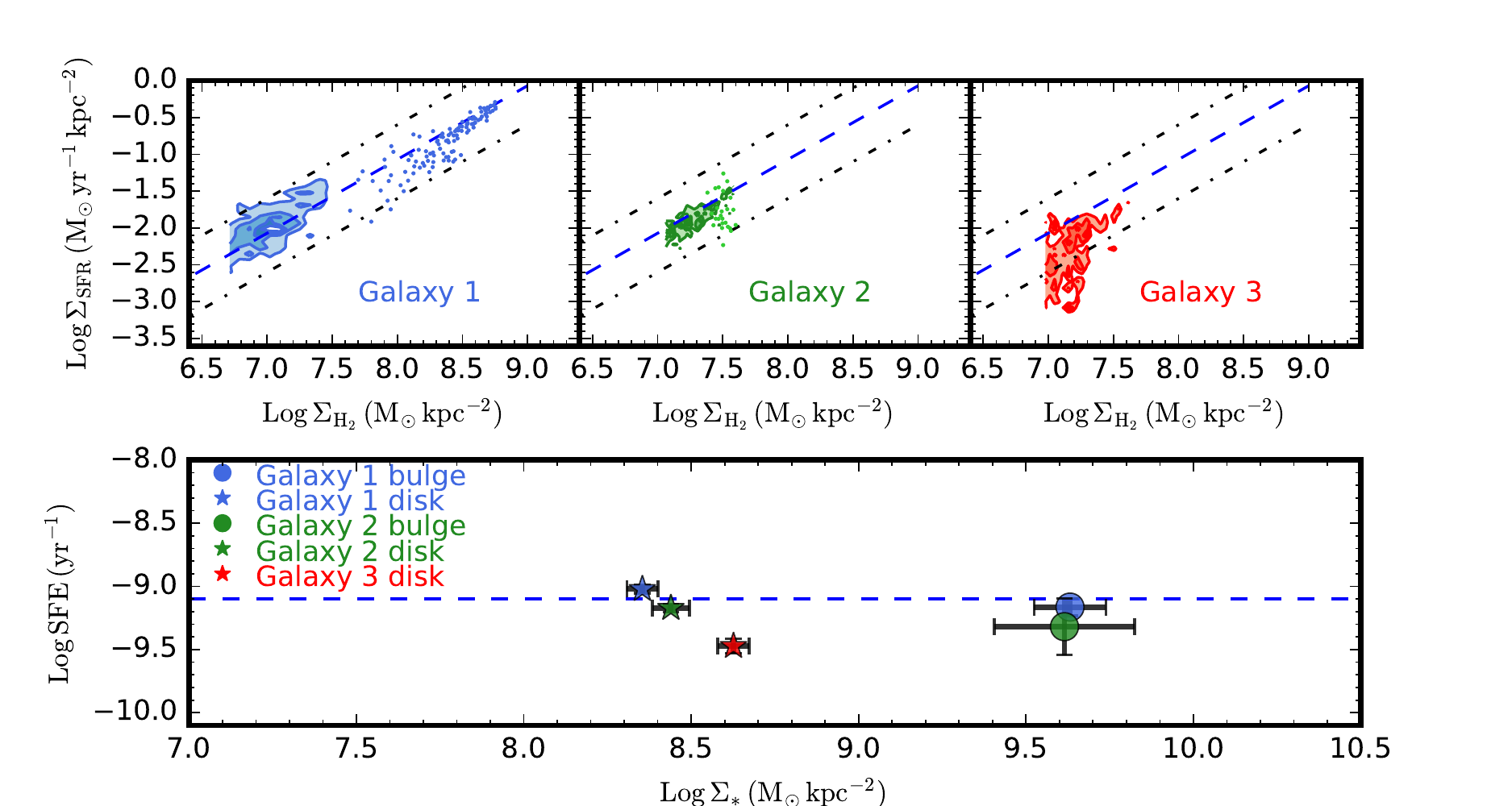}
\caption{Top panels: The relation between the SFR surface density and the gas surface density for Galaxy 1, Galaxy 2, and Galaxy 3 (from left to right). The dashed line corresponds to a constant SFE of $10^{-9.1}$ yr$^{-1}$, the averaged result from the HERACLES sample \citep{ler08}. while the upper and lower doted-dashed lines correspond to constant gas fractions of $10^{-8.6}$ and $10^{-9.6}$ yr$^{-1}$, respectively. Bottom panel: The star formation efficiency as a function of stellar mass surface density. The dashed line shows a constant SFE of $10^{-9.1}$ yr$^{-1}$. Colors, symbols, and the method used to compute the uncertainties are same as those in Figure \ref{fig:gas-sm}.  \label{fig:sfr-h2}}
\end{figure*}

\begin{figure*}
\centering
\includegraphics[angle=0,width=0.9\textwidth]{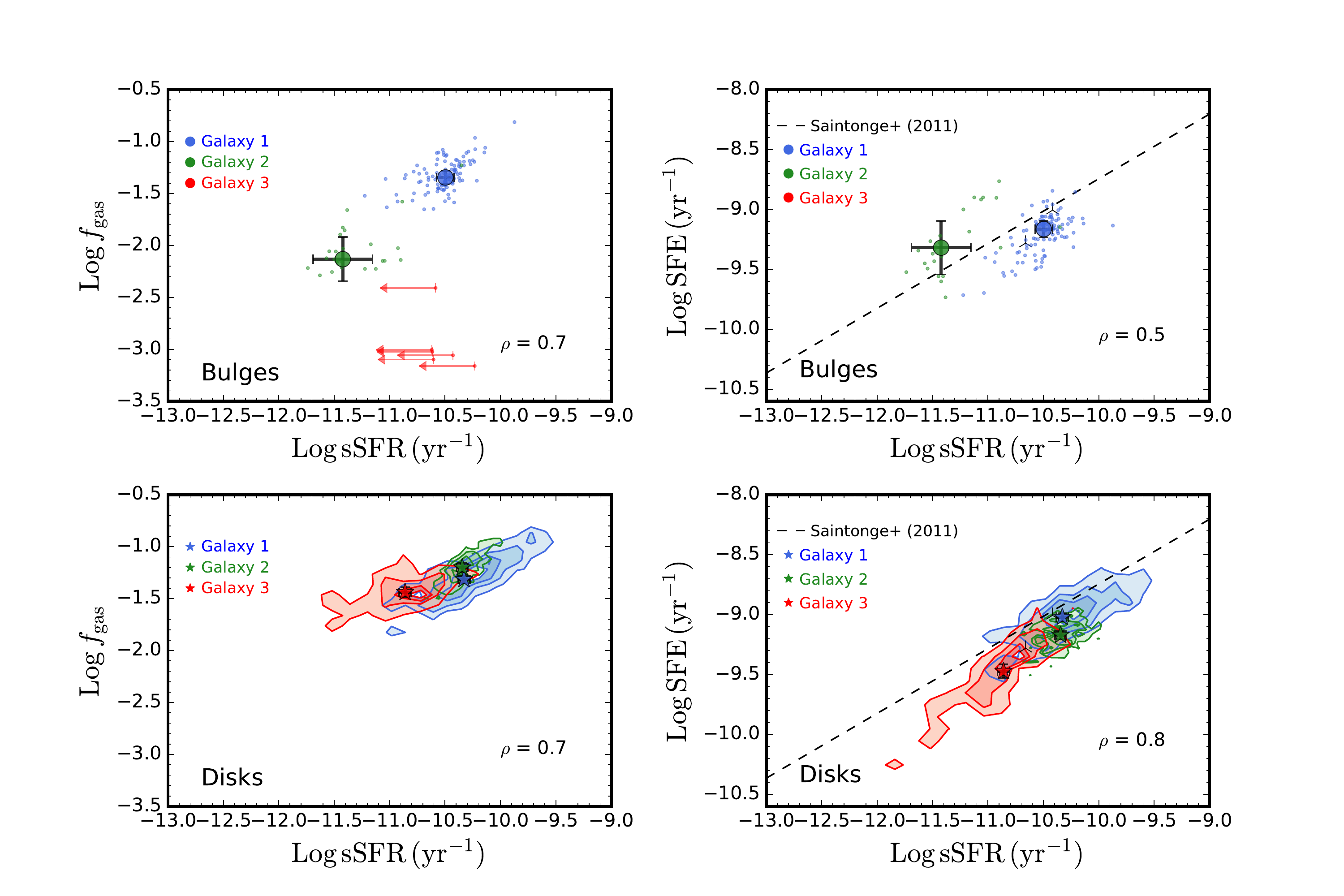}
\caption{Left panels: The gas fraction as a function of specific star formation rate. Right panels: The star formation efficiency as a function of specific star formation rate. The dashed lines show the COLD GASS results based on a sample of galaxies with secure CO detections (see Table 1 of \citet{sai11}). The small dots in the upper two panels represent spaxels located in the bulges. In the bottom two panels, the contours in the lower two panels represent the distributions in the disks. The solid circles and stars show the median values of the bulge and disk regions, respectively. The red arrows in the top-left panel indicate the upper limits for Galaxy 3. Colors, symbols, and the method used to compute the uncertainties are same as those in Figure \ref{fig:gas-sm}. The Spearman's correlation coefficient $\rho$ is also shown in the lower-right corner of each panel.  \label{fig:gas-sSFR}}
\end{figure*}

To separate the bulge and disk regions in our data, we perform the two-component fitting using \texttt{GALFIT} \citet{pen02,pen10} in the SDSS $r-$band images. For the bulge and disk components, we fix the Sersic index to be $n = 4$ and $n = 1$, respectively, when fitting other parameters. The effective radii ($R_e$) are determined to be 2.77\arcsec, 0.7\arcsec, and 1.23\arcsec for Galaxies 1, 2, and 3, respectively. Once we obtain $R_e$ of the bulge, we compute the observed effective radius ($R_e^{obs}$) by convolving it with the PSF size of both MaNGA and ALMA beams ($\sim$ 2.5"). We define the "bulge" region to be $r <  R_e^{obs}$, and in order to mitigate contamination from any overlap region we define the "disk" region to be $r > $ 2$\times R_e^{obs}$. These regions are indicated by white circles in Figure \ref{fig:map}.  

\subsection{Resolved \sigSFR~vs. \sigsm~relation}

Recently, it has been found that kpc-scale \sigSFR~ traces well with the underlying \sigsm~ for star-forming galaxies \citep{san13,can16,hsi17}. This relation may be responsible for the observed tight correlation between the global SFR and \sm. In the right panel of Figure \ref{fig:sfr-sm}, we show the kpc-scale \sigSFR~ vs. \sigsm~relation for the 3 green valley galaxies. The data points from the bulge and disk regions are shown in solid circles and contours, respectively. The dotted line represents the best-fit of the resolved main sequence relation obtained for the MaNGA star-forming population \citep{hsi17}: log($\frac{\rm SFR}{\rm M_{\odot} yr^{-1}}$) = $-$10.33 + log($\frac{\rm M_{*}}{\rm M_{\odot}})$.   
For Galaxy 1, the disk almost lies on the resolved main sequence while the bulge is only slightly below the line. On the other hand, the bulge of Galaxy 2 shows significant departure from the resolved main sequence. We note that the central \ha~emission of Galaxy 3 is dominated by a broad emission lines associated with an AGN, and therefore the \sigSFR~from \ha~ is an upper limit in the central part of this galaxy. In spite of this, it is clear that both the bulge and disk regions of Galaxy 3 are systematically below the resolved main-sequence relation. 
There is also a trend that the disk sSFR declines from Galaxy 1 to Galaxy 3, following a similar behavior of the global sSFR. Assuming galaxies evolve with declining sSFR, our results would indicate that the sSFR in bulge departs from the resolved main sequence first, followed by the disk as the global sSFR decreases \footnote{It is not necessarily true that these three galaxies form an evolutionary sequence, in particular because Galaxy 2 is less massive than the other two. In addition, the sSFR of galaxies may not monotonically declines with time as the SFR can be re-ignited by various processes during the life time of galaxies.}.

\subsection{Gas fraction, star formation efficieny, and specific star formation rate} 
Figure \ref{fig:map} shows the optical image, \sigSFR, \sigh2, star formation efficiency (SFE; defined as SFR/\Mh2), and the gas fraction (\fgas; defined as \sigh2/(\sigsm+\sigh2))  for the three green valley galaxies. We can see that the spatial distributions of these quantities are diverse among the three galaxies. For example, both \sigSFR~ and \sigh2~peak in the central part of Galaxy 1, whereas \sigh2~is more evenly distributed in Galaxy 2 and greater in the outskirts in Galaxy 3. Interestingly, all 3 galaxies show increasing \fgas~ with radius. We note that MaNGA achieves nearly uniform sensitivity across its IFU bundles, which is sufficient to detect the continuum at high S/N in the outskirts of these objects. The observed increases in gas fraction is, therefore, not driven by low S/N in the outskirts of galaxies. To quantify these differences, we next compare the relations among various quantities. The upper panels of Figure \ref{fig:gas-sm} shows the \sigh2~
versus \sigsm~ relation. For Galaxy 1, \sigh2~scales with \sigsm~ for both 
bulge and disk regions. On the other hand, for Galaxies 2 and 3, \sigh2 
is quite uniform across the bulges, despite the fact that the \sigh2~ correlates with \sigsm~ in 
disks. The lower panel shows the median gas fraction as a function of \sigsm~ in bulges (circles) and disks (stars). The gas fraction in the bulges varies significantly among the 3 galaxies by 1.6 dex, being lower toward the Galaxy 3. On the other hand, the gas fractions in disks are comparable in the three cases, although slightly lower in Galaxy 3. Except for Galaxy 1, \fgas~ in the bulges is significantly lower than in the disks for the other two galaxies. 

Next, we explore the relation between the SFR surface density and the gas surface density, the so-called `Kennicutt-Schmidt' relation \citep{ken98}, shown in the upper panels of Figure \ref{fig:sfr-h2}. Only spaxels with S/N (CO) > 2 are displayed. The SFE versus \sigsm~ is shown in the lower panel\footnote{Here we only consider the molecular gas, not the HI gas mass}. Except for the Galaxy 3 whose central
\ha~ emission is contaminated by the Broad Line Region associated with the AGN, the SFEs of other two bulges are at a similar level and are moderately lower compared to their corresponding disks. Similarly, the disk regions show a wider spread in the resolved SFE. Even though the SFEs are similar in some regions among the three galaxies, their median values systematically decline from 
Galaxies 1 to 3 by a factor of 3. Since two of our targets also have single-dish CO observations, we can  compare the resolved SFE with the global SFE measurements as listed in Table \ref{tab:property}. The global SFE  is in good agreement with the resolved SFE for Galaxy 1. On the other hand, the global SFE is close to the lower end of the  resolved SFE distribution for Galaxy 2. We note that the later is caused by the factor of 1.9 excess in the total CO (1-0) flux estimated by the COLD GASS single dish measurement compared to the integrated ALMA flux (see Sec.2.2).

To address the relative importance in controlling the sSFR between gas fraction and SFE, we plot \fgas~ and SFE against sSFR and compute the Spearman's correlation coefficient $\rho$ as shown in Figure \ref{fig:gas-sSFR}. The data points associated with the bulge regions in Galaxy 3 are excluded in this analysis given that we cannot measure the SFR directly due to the AGN contamination. 
For bulges, the relation between \fgas~ an sSFR is stronger than that between SFE and sSFR as indicated by the $\rho$ values, suggesting that the sSFR of bulges is mainly controlled by \fgas. On the other hand, it is observed that both local \fgas~ and SFE correlate with local sSFR in disks, and the local relations are common among the 3 disks. For comparison, we also plot the global SFE vs. sSFR relations of the COLD GASS sample for galaxies with secure CO detections \citep{sai11} in Figure \ref{fig:gas-sSFR} after correcting for the differences in the adopted IMF and $\alpha_{\mathrm{CO}}$. Our data points in the disk regions are systematically below the best-fit lines of the COLD GASS sample. This discrepancy may come from the fact that these two samples are averaged over different physical scales. The spatially resolved observations tend to sample CO bright regions, resulting in lower SFE  than the global averages. An alternative explanation is that green valley galaxies may form a different correlation from the main sequence. Observations covering a wider range of galaxy populations are needed to conclude whether the observed SFE vs. sSFR relation is universal.

\section{Discussion}
The above analyses suggest that \fgas~ in bulges declines dramatically from Galaxy 1 to Galaxy 3, while keeping a similar level of SFE compared to the disks. On the other hand, there is a significant decrease in the disk SFE when the global sSFR of galaxies drops. To first order, the sSFR $\sim$ \fgas~$\times$ SFE. If we assume that green valley galaxies evolve with a declining global sSFR, a plausible scenario is that the bulge first quenched due to the reduction on the cold gas available, followed by a subsequent quenching in the disk because of the decrease in both the SFE and \fgas. This is consistent with the inside-out quenching scenario in which the star formation is ceases in bulges first. 

The physical cause of the inside-out quenching, however, remains unanswered. 
Using the sample drawn from the HERA CO-Line Extragalactic Survey (HERACLES; Leroy et al. 2008), \citet{hua15} found that the gas depletion time is shorter for the bulge than for the disk. Although the galaxies used in their works is part of an HI-selected sample, which is mainly composed of normal star-forming galaxies on the main sequence, their result indicates that the greater SFE of the bulge may be responsible in reducing the amount of cold gas for the bulge, leading to the quenching of star formation if there is no further gas supply for the bulge. The observation of our three green valley galaxies does not seem to share the same trend, in particular that the bulge SFEs of green valley galaxies are similar or even lower than what is observed in their disks, similar to the findings by \citep{fis13} based on a combined sample of the BIMA SONG \citep{hel03} , CARMA STING \citep{rah12}, and PdBI NUGA \citep{gar03}
surveys. Our Galaxy 1, which is closest to the main sequence, shows comparable SFE and \fgas~between its bulge and disk, suggesting that star formation alone cannot explain the faster reduction of the cold molecular gas in the bulges; some other processes are required to efficiently reduce or remove the cold molecular gas in the central parts of galaxies when galaxies migrate to the quiescent population.

One of the commonly accepted pictures refers to the so-called AGN feedback which heats up or expels the surrounding gases, preventing galaxies from subsequent star formation, particularly for massive galaxies. This scenario is supported by the observation of low gas fraction in AGN host galaxies \citep[e.g.,][]{bru15,kak17}, as well as the AGN-driven molecular gas outflow \citep{cic14,fer15}. In addition, it has been reported that AGN hosts preferentially lie in the green valley or below the main sequence \citep{nan07,sil08,sal07,ell16,smi16}, suggesting that AGN could drive the transition from the star-forming to the quiescent phases. Recently, a spatially-resolved star formation rate study using MaNGA galaxies also finds that the resolved sSFR of unbarred AGN hosts is below the resolved main-sequence (L. Bing et al. in prep.) across the entire galaxies, similar to the three cases presented in this work. As noted earlier, one of our three green valley galaxies, Galaxy 3, shows broad-line features and hence potentially hosts an AGN. It is consistent with the AGN feedback framework that the presence of AGN diminishes the available cold gas in the bulge and even in the disks. On the other hand, although the morphological quenching \citep{mar09} also predicts low SFE in the disks that are stabilized against gas fragmentation due to the presence of massive bulges, it may not be relevant to the three systems discussed in this work since morphological quenching is only effective in bulge dominated systems, unlike our green valley galaxies.

\section{CONCLUSIONS}\label{sec:conclusion}
We have observed three MaNGA-selected green valley galaxies with ALMA CO (1-0) to study the role of gas in star formation quenching. The three galaxies are referred to as Galaxies 1, 2, and 3 according to their separation from the main sequence on the global sSFR and \sm~relation (1: closest; 3: farthest). Specifically, we study the relations among  sSFR, SFE, and \fgas~on kpc scales. Our results can be summarized below:

1. The resolved MaNGA data show that the disk sSFR declines with the decreasing global sSFR. There is an indication that the bulge departs from the resolved main sequence first, followed by the disk as the global sSFR declines.

2. For Galaxies 2 and 3, which are clearly below the star-forming main sequence, the gas fraction in the bulges is lower compared to that in the disks. The gas fraction in the bulges drops by 1.6 dex from Galaxy 1 to Galaxy 3.

3. The SFE in the bulge is moderately lower than that in the disk for Galaxy 1 and Galaxy 2. In addition, the SFE of disks decreases from Galaxy 1 to Galaxy 3.

4. The resolved sSFR is found to correlate with both \fgas~and SFE. However, the sSFR of bulges have stronger dependence on \fgas. On the other hand, the resolved sSFR in disks are sensitive to both \fgas~and SFE.\\

Our results suggest that the \fgas~ is the dominant factor determining the sSFR of bulges, while the sSFR of disks declines because of the drop in both SFE and \fgas~ when the global sSFR declines. Assuming the three galaxies represent a sequence of transitional stages, our results would favour an inside-out quenching -- the SF is ceased in the bulge first because of the lack of available cold gas, followed by the quenching in the disk due to subsequent decline in SFE as well as in \fgas.  Our results fit into the AGN feedback scenario in which the AGN activity may heat up or eject the cold gas out, resulting a reduction of available cold gas to fuel the star formation in the bulges (and possible in the disks), although it remains unclear what drives the declination of SFE in the disks when galaxies move away from the main sequence. However, such a evolution sequence may be oversimplified as it has been shown that green valley galaxies can have diversity in terms of their quenching time scales, suggesting different pathways of star formation quenching \citep{sch14,sme15}. Moreover, galaxies may be rejuvenated if there is fresh gas accreted when experiencing minor mergers or galaxy interactions \citep{tho10,hai15,lac16} and hence may not evolve monotonically with a decreasing global sSFR.  Detailed stellar population analyses regarding the star formation and stellar mass assembling histories \citep[e.g.,][]{iba16} together with resolved gas observations for a larger sample of green valley galaxies is required to confirm the picture presented in this work.

\acknowledgments

We thank the anonymous referee for valuable suggestions that significantly improve the contents of this paper. We thank C. Maraston, E.  Emsellem, M. Cappellari, and A. Aragon-salamanca for helpful suggestions. The work is supported by the Ministry of Science \& Technology of Taiwan
under the grant MOST 103-2112-M-001-031-MY3. R.M. and F.B. acknowledge support by the UK Science and Technology Facilities Council (STFC).
R.M. acknowledges ERC Advanced Grant 695671 `QUENCH'. 

This paper makes use of the following ALMA data: ADS/JAO.ALMA\#2015.1.01225.S. ALMA is a partnership of ESO (representing its member states), NSF (USA) and NINS (Japan), together with NRC (Canada), NSC and ASIAA (Taiwan), and KASI (Republic of Korea), in cooperation with the Republic of Chile. The Joint ALMA Observatory is operated by ESO, AUI/NRAO and NAOJ. This project also makes use of the MaNGA-Pipe3D dataproducts. We thank the IA-UNAM MaNGA team for creating it, and the ConaCyt-180125 project for supporting them. The Green Bank Observatory is a facility of the National Science Foundation. This work used data from project AGBT17A\_012: `HI-MaNGA: HI Followup of MaNGA galaxies', PI Karen L. Masters.

Funding for the Sloan Digital Sky Survey IV has been
provided by the Alfred P. Sloan Foundation, the U.S.
Department of Energy Office of Science, and the Participating Institutions. SDSS-IV acknowledges support
and resources from the Center for High-Performance
Computing at the University of Utah. The SDSS web
site is www.sdss.org. SDSS-IV is managed by the Astrophysical Research Consortium for the Participating
Institutions of the SDSS Collaboration including the
Brazilian Participation Group, the Carnegie Institution
for Science, Carnegie Mellon University, the Chilean
Participation Group, the French Participation Group,
Harvard-Smithsonian Center for Astrophysics, Instituto
de Astrof\'isica de Canarias, The Johns Hopkins University, Kavli Institute for the Physics and Mathematics of the Universe (IPMU) / University of Tokyo, Lawrence
Berkeley National Laboratory, Leibniz Institut f\"ur Astrophysik Potsdam (AIP), Max-Planck-Institut f\"ur Astronomie (MPIA Heidelberg), Max-Planck-Institut f\"ur
Astrophysik (MPA Garching), Max-Planck-Institut f\"ur
Extraterrestrische Physik (MPE), National Astronomical Observatory of China, New Mexico State University,
New York University, University of Notre Dame, Observat\'ario Nacional / MCTI, The Ohio State University,
Pennsylvania State University, Shanghai Astronomical
Observatory, United Kingdom Participation Group, Universidad Nacional Aut\'onoma de M\'exico, University of
Arizona, University of Colorado Boulder, University of
Oxford, University of Portsmouth, University of Utah,
University of Virginia, University of Washington, University of Wisconsin, Vanderbilt University, and Yale University.


\end{document}